\documentclass[12pt]{article}%
\usepackage{amsmath,latexsym}
\usepackage{graphicx}
\usepackage{amsmath}
\usepackage{amsfonts}
\usepackage{amssymb}%
\begin{document}

\title{{Periodic signature change in spacetimes of
    embedding class one}}
   \author{
Peter K. F. Kuhfittig*\\
\footnote{kuhfitti@msoe.edu}
 \small Department of Mathematics, Milwaukee School of
Engineering,\\
\small Milwaukee, Wisconsin 53202-3109, USA}

\date{}
 \maketitle

\begin{abstract}\noindent
The idea of an oscillating Universe has
remained a topic of interest even after
the discovery of dark energy.  This paper
confirms this idea by means of another
well-established theory in general
relativity, the embedding of curved
spacetimes in higher-dimensional flat
spacetimes: an $n$-dimensional Riemannian
space is said to be of embedding class
$m$ if $m+n$ is the lowest dimension $d$ 
of the flat space in which the given space
can be embedded; here $d=\frac{1}{2}n(n-1)$.
So a four-dimensional Riemannian space
is of class two sine it can be embedded 
in a six-dimensional flat space.  A
line element of class two can be reduced
to a line element of class one by a
suitable coordinate transformation.  The
extra dimension can be either spacelike
or timelike, leading to accelerating and
decelerating expansions, respectively.
Accordingly, it is proposed in this paper
that the free parameter occurring in the
transformation be a periodic function of
time.  The result is a mathematical model
that can be interpreted as a periodic
change in the signature of the embedding
space. This signature change may be the
best model for an oscillating Universe and
complements various models proposed in
the literature.
\\
\end{abstract}

\section{Introduction}

The idea of a signature change in general
relativity is not new: the Schwarzschild
line element

\begin{equation}\label{E:line1}
  ds^2=-\left(1-\frac{2M}{r}\right)dt^{2
  }+\frac{dr^2}{1-2M/r}
+r^{2}(d\theta^{2}+\text{sin}^{2}\theta\,
d\phi^{2})
\end{equation}
implies that space and time are interchanged
when crossing the event horizon of a black
hole.  Signature-changing events can also
play an important role in quantum gravity in
part because a quantum field defined on a
manifold may contain regions of different
signatures \cite{WWV10}.  Modifications of
Einstein's theory have led to similar
considerations, an example of which is $f(R,T)$
gravity discussed in Ref. \cite{STP18}.  Here
$R$ is the Ricci scalar and $T$ is the trace
of the energy-momentum tensor.  Given the
current accelerated expansion, a transition
from a previous decelerated phase suggests a
time-varying deceleration parameter.  According
to Ref. \cite{STP18}, the evolving deceleration
parameter displays a signature-flipping
behavior.  The outcome is an oscillating
Universe.  A similar argument is presented
in Ref. \cite{LMT21}, based on $f(Q,T)$
teleparallel gravity.  Refs. \cite{aU08} and
\cite{PPG} discuss the concept of an
oscillatory Universe from a different
perspective, a dynamic dark-energy equation
of state. This process would continue
indefinitely without violating any known
conservation laws.

Totally different scenarios are considered in
Refs. \cite{MSV08} and \cite{BB20}.  The
discussion in Ref. \cite{MSV08} involves a
brane-world model in a five-dimensional
anti-de Sitter space.  It is argued that the
present accelerated expansion may simply be
an indication that our brane world has
undergone a signature change.  Ref. \cite
{BB20} returns to quantum gravity via
covariant models in loop quantum gravity.
It is asserted that these models generically
imply dynamical signature changes at high
densities.

Other mathematical models of interest to
us involve various embedding theorems.
These have a long history in the general
theory of relativity, aided in large part
by Campbell's theorem \cite{jC26}.  According
to Ref. \cite{pW15}, the five-dimensional
field equations in terms of the Ricci tensor,
i.e., $R_{AB}=0$, $A,B=0,1,2,3,4$, help
explain the origin of matter.  The reason
is that the vacuum field equations in five
dimensions yield the usual Einstein field
equations \emph{with matter}, called the
induced-matter theory \cite{WP92, SW03}:
what we perceive as matter can be viewed
as the impingement of the extra dimension
onto our spacetime.  Moreover, the extra
dimension can be either spacelike or
timelike.  According to Ref. \cite{pW11},
the particle-wave duality can in principle
be solved because the five-dimensional
dynamics has two modes, depending on
whether the extra dimension is spacelike
or timelike.  So it is conceivable that
the five-dimensional relativity theory
could lead to a unification of general
relativity and quantum field theory.

To become a useful mathematical model, we
need to go beyond a single extra dimension.
This requires the concept of embedding: an
$n$-dimensional Riemannian space is said to
be of embedding class $m$ if $m+n$ is the
lowest dimension $d$ of the flat space in which
the given space can be embedded; here
$d=\frac{1}{2}n(n-1)$. So a four-dimensional
Riemannian space is of class two since it 
can be embedded in a six-dimensional
flat space.  Moreover, a line element of
class two can be reduced to a line element
of class one by a suitable coordinate
transformation \cite{MDRK, MRG17, MM17, MG17}.
As noted above, the extra dimension can be
spacelike or timelike.  Not only is the
result a powerful mathematical model, it is
consistent with observation since the extra
dimension could not be directly observed.

The mathematical model proposed contains a
free parameter $K=K(t)$, a periodic function
of time that causes a periodic change in the
signature of the embedding space.  It is
shown in this paper that a positive value
of $K$ can account for the accelerating
expansion, while a negative value
corresponds to a decelerating phase.  The
resulting oscillatory behavior can thereby
be attributed to a signature change in the
embedding space, complementing the above
mathematical models.

This paper is organized as follows: Sec.
\ref{S:embedding} describes the embedding
of a curved spacetime in a flat spacetime
with an extra spacelike or timelike dimension.
Sec. \ref{S:K(t)} discusses the free parameter
$K$, now assumed to be a periodic function
of time, to provide an effective model for a
signature change. The result is the oscillatory
behavior discussed in Sec. \ref{S:period}.
In Sec. \ref{S:summary} we conclude.

\section{The embedding}\label{S:embedding}
\noindent
Following Ref. \cite{MDRK}, we begin with the
spherically symmetric line element
\begin{equation}\label{E:line2}
ds^{2}=-e^{\nu(r)}dt^{2}+e^{\lambda(r)}dr^{2}
+r^{2}\left(d\theta^{2}+\sin^{2}\theta \,d\phi^{2}
\right),
\end{equation}
using units in which $c=G=1$.  We also assume
asymptotic flatness: $e^{\nu(r)}\rightarrow 1$
and $e^{\lambda(r)}\rightarrow 1$ as
$r\rightarrow \infty$.  As already noted, this
metric of class two can be reduced to a metric
of class one by a suitable transformation of
coordinates.  Since the extra dimension in the
embedding space can be either spacelike or
timelike, the two cases will be taken up
separately.

\subsection{An extra spacelike dimension}
In this subsection, we will consider the
following five-dimensional embedding space:
\begin{equation}\label{E:line3}
ds^{2}=-\left(dz^1\right)^2+\left(dz^2\right)^2
+\left(dz^3\right)^2+\left(dz^4\right)^2
+\left(dz^5\right)^2.
\end{equation}
The transformation is given by
$z^1=\sqrt{K}\,e^{\frac{\nu}{2}}
 \,\text{sinh}{\frac{t}{\sqrt{K}}}$,
  $z^2=\sqrt{K}
 \,e^{\frac{\nu}{2}}\,\text{cosh}{\frac{t}{\sqrt{K}}}$,
 $z^3=r\,\text{sin}\,\theta\,\text{cos}\,\phi$, $z^4=
 r\,\text{sin}\,\theta\,\text{sin}\,\phi$,
 and
 $z^5=r\,\text{cos}\,\theta$, also discussed in
 Ref. \cite{KG18}.  (Some of the steps are
 repeated here for later reference.)  The
 differentials of these components are
\begin{equation}\label{E:diff1}
dz^1=\sqrt{K}\,e^{\frac{\nu}{2}}\,\frac{\nu'}{2}\,
\text{sinh}{\frac{t}{\sqrt{K}}}\,dr + e^{\frac{\nu}{2}}\,
\text{cosh}{\frac{t}{\sqrt{K}}}\,dt,
\end{equation}
\begin{equation}\label{E:diff2}
dz^2=\sqrt{K}\,e^{\frac{\nu}{2}}\,\frac{\nu'}{2}\,
\text{cosh}{\frac{t}{\sqrt{K}}}\,dr + e^{\frac{\nu}{2}}\,
\text{sinh}{\frac{t}{\sqrt{K}}}\,dt,
\end{equation}
\begin{equation}
dz^3=\text{sin}\,\theta\,\text{cos}\,\phi\,dr + r\,
\text{cos}\,\theta\,\text{cos}\,\phi\,
d\theta\,-r\,\text{sin}\,\theta\,\text{sin}\,\phi\,d\phi,
\end{equation}

\begin{equation}
dz^4=\text{sin}\,\theta\,\text{sin}\,\phi\,dr + r\,
\text{cos}\,\theta\,\text{sin}\,\phi\,
d\theta\,+r\,\text{sin}\,\theta\,\text{cos}\,\phi\,d\phi,
\end{equation}
and
\begin{equation}
dz^5=\text{cos}\,\theta\,dr\, - r\,\text{sin}\,\theta\,d\theta.
\end{equation} To facilitate the substitution
into Eq. (\ref{E:line2}), we first obtain the expressions
for $-\left(dz^1\right)^2+\left(dz^2\right)^2$ and for
$\left(dz^3\right)^2+\left(dz^4\right)^2
+\left(dz^5\right)^2$:
\begin{equation*}
  -\left(dz^1\right)^2+\left(dz^2\right)^2=
  -e^{\nu}dt^{2}+\,\frac{1}{4}K\,e^{\nu}\,
(\nu')^2\,\,dr^{2}
\end{equation*}
and
\begin{equation*}
  \left(dz^3\right)^2+\left(dz^4\right)^2
+\left(dz^5\right)^2=dr^2+r^{2}\left(d\theta^{2}
+\sin^{2}\theta\, d\phi^{2} \right).
\end{equation*}
The substitution yields
\begin{equation}\label{E:line4}
ds^{2}=-e^{\nu}dt^{2}+\left(\,1+\frac{1}{4}K\,e^{\nu}\,
(\nu')^2\,\right)\,dr^{2}+r^{2}\left(d\theta^{2}
+\sin^{2}\theta\, d\phi^{2} \right).
\end{equation}
Metric (\ref{E:line4}) is therefore equivalent to
metric (\ref{E:line2}) if
\begin{equation}\label{E:lambda}
e^{\lambda}=1+\frac{1}{4}K\,e^{\nu}\,(\nu')^2,
\end{equation}
where $K>0$ is a free parameter.  The result is
a metric of embedding class one.  Eq. (\ref{E:lambda})
can also be obtained from the Karmarkar condition
\cite{kK48}:
\begin{equation}
   R_{1414}=\frac{R_{1212}R_{3434}+R_{1224}R_{1334}}
   {R_{2323}},\quad R_{2323}\neq 0.
\end{equation}
In fact, Eq. (\ref{E:lambda}) is a solution of
the differential equation
\begin{equation}
   \frac{\nu'\lambda'}{1-e^{\lambda}}=
   \nu'\lambda'-2\nu''-(\nu')^2,
\end{equation}
which is readily solved by separation of
variables.  So $K$ is actually an arbitrary
constant of integration.

\subsection{An extra timelike dimension}

Here we consider the following embedding space:
\begin{equation}
ds^{2}=-\left(dz^1\right)^2-\left(dz^2\right)^2
+\left(dz^3\right)^2+\left(dz^4\right)^2
+\left(dz^5\right)^2.
\end{equation}
It is shown in Ref. \cite{KG18} that the
coordinate transformation is
$z^1=\sqrt{K}\,e^{\frac{\nu}{2}}
 \,\text{sin}{\frac{t}{\sqrt{K}}}$,
  $z^2=\sqrt{K}\,e^{\frac{\nu}{2}}\,\text{cos}
  {\frac{t}{\sqrt{K}}}$,
  $z^3=r\,\text{sin}\,\theta\,\text{cos}\,\phi$,
  $z^4=r\,\text{sin}\,\theta\,\text{sin}\,\phi$,
 and
 $z^5=r\,\text{cos}\,\theta$.

 The substitution yields
\begin{equation}
ds^{2}=-e^{\nu}dt^{2}+\left(\,1-\frac{1}{4}K\,e^{\nu}\,
(\nu')^2\,\right)\,dr^{2}+r^{2}\left(d\theta^{2}
+\sin^{2}\theta\, d\phi^{2} \right).
\end{equation}
So
\begin{equation}\label{E:lambda2}
e^{\lambda}=1-\frac{1}{4}K\,e^{\nu}\,(\nu')^2,
\end{equation}
where $K$ is a constant of integration, as before.

\section{A variable parameter: $K=K(t)$}
   \label{S:K(t)}
The similarity between Eqs. (\ref{E:lambda})
and (\ref{E:lambda2}) suggests a direct
connection that can be attributed to a
signature change.  In Eq. (\ref{E:line2}),
both $\nu$ and $\lambda$ are functions of
$r$.  So the parameter $K$ in Eq.
(\ref{E:lambda}) could be a function of
time, which is also true of the parameter
$K$ in Eq. (\ref{E:lambda2}), i.e., $K=K(t)$.
To see the significance, suppose that $K(t)$
is a periodic real-valued function that is
symmetric with respect to the $t$-axis,
similar to, for example, the function $f(t)=
A\,\text{sin}\,\omega t$; so $|K(t)|=|-K(t)|$.
To see the effect on the above transformation,
let us replace $K$ by $-K$ in Eqs.
(\ref{E:diff1}) and (\ref{E:diff2}):
\begin{equation}
dz^1=\sqrt{-K}\,e^{\frac{\nu}{2}}\,\frac{\nu'}{2}\,
\text{sinh}{\frac{t}{\sqrt{-K}}}\,dr + e^{\frac{\nu}{2}}\,
\text{cosh}{\frac{t}{\sqrt{-K}}}\,dt
\end{equation}
and
\begin{equation}
dz^2=\sqrt{-K}\,e^{\frac{\nu}{2}}\,\frac{\nu'}{2}\,
\text{cosh}{\frac{t}{\sqrt{-K}}}\,dr + e^{\frac{\nu}{2}}\,
\text{sinh}{\frac{t}{\sqrt{-K}}}\,dt.
\end{equation}
Making use of the identities
  \[\text{sinh}(ix)=i\,\text{sin}\, x\quad
  \text{and}\quad \text{cosh}(ix)=\text{cos}\, x,\]
we obtain
\begin{equation}
dz^1=\sqrt{K}\,e^{\frac{\nu}{2}}\,\frac{\nu'}{2}\,
\text{sin}{\frac{t}{\sqrt{K}}}\,dr + e^{\frac{\nu}{2}}\,
\text{cos}{\frac{t}{\sqrt{K}}}\,dt
\end{equation}
and
\begin{equation}
dz^2=i\sqrt{K}\,e^{\frac{\nu}{2}}\,\frac{\nu'}{2}\,
\text{cos}{\frac{t}{\sqrt{K}}}\,dr -i e^{\frac{\nu}{2}}\,
\text{sin}{\frac{t}{\sqrt{K}}}\,dt.
\end{equation}
The result is
\begin{equation*}
  -\left(dz^1\right)^2+\left(dz^2\right)^2=
  -e^{\nu}dt^{2}-\,\frac{1}{4}K\,e^{\nu}\,
(\nu')^2\,\,dr^{2}.
\end{equation*}
Since
\begin{equation*}
  \left(dz^3\right)^2+\left(dz^4\right)^2
+\left(dz^5\right)
\end{equation*}
remains the same, we obtain the line element
\begin{equation}
ds^{2}=-e^{\nu}dt^{2}+\left(\,1-\frac{1}{4}K\,e^{\nu}\,
(\nu')^2\,\right)\,dr^{2}+r^{2}\left(d\theta^{2}
+\sin^{2}\theta\, d\phi^{2} \right).
\end{equation}
It follows that
\begin{equation}
e^{\lambda}=1-\frac{1}{4}K\,e^{\nu}\,(\nu')^2,
\end{equation}
in agreement with Eq. (\ref{E:lambda2}).
So changing the sign of the parameter $K$
has the same effect as changing the extra
spacelike dimension in the embedding space
to a timelike dimension.  The periodicity
of $K(t)$ therefore results in a periodic
signature change in the embedding space
and suggests an oscillating behavior.
The periodicity of $K$ produces a
mathematical model that is consistent
with the equation of state $p=\omega\rho$
discussed in Ref. \cite{PPG}, where
$\omega(t)=\omega_0+\omega_1(t\dot{H}/H)$.
Here $H=H(t)$ is the time varying Hubble
parameter resulting in both expanding and
contracting epochs of the Universe.
Similarly, Ref. \cite{bF06} discusses a
form of dark energy called quintom
having an oscillating equation of state,
which, in turn, leads to oscillations of
the Hubble parameter and a recurring
Universe.

\section{The oscillating Universe}
    \label{S:period}

Returning to line element (\ref{E:line2}), let
us first list the Einstein field equations:
\begin{equation}\label{E:E1}
8\pi \rho=e^{-\lambda}
\left(\frac{\lambda^\prime}{r} - \frac{1}{r^2}
\right)+\frac{1}{r^2},
\end{equation}
\begin{equation}\label{E:E2}
8\pi p=e^{-\lambda}
\left(\frac{1}{r^2}+\frac{\nu^\prime}{r}\right)
-\frac{1}{r^2},
\end{equation}
and
\begin{equation}\label{E:E3}
8\pi p_t=
\frac{1}{2} e^{-\lambda} \left[\frac{1}{2}(\nu^\prime)^2+
\nu^{\prime\prime} -\frac{1}{2}\lambda^\prime\nu^\prime +
\frac{1}{r}({\nu^\prime- \lambda^\prime})\right].
\end{equation}
Next, we state the
Friedmann-Lema\^{i}tre-Robertson-Walker (FLRW) model
in the usual four dimensions \cite{rW84}:
\begin{equation}
ds^{2}=-dt^{2}+a^2(t)\left[\frac{dr^2}{1-kr^2}
+r^{2}\left(d\theta^{2}+\sin^{2}\theta \,d\phi^{2}
\right)\right],
\end{equation}
where $a^2(t)$ is a scale factor.  In this
paper we also make use of the Friedmann equation
\begin{equation}\label{E:Friedmann}
  \frac{\overset{..}{a}(t)}{a(t)}=
  -\frac{4\pi}{3}(\rho +3p),
\end{equation}
again using units in which $c=G=1$.

Before continuing, we need one more important
observation: in the outer region of a galactic
halo, we have \cite{Nandi}
\begin{equation}
ds^{2}=-B_0r^ldt^{2}+e^{\lambda(r)}dr^{2}
+r^{2}\left(d\theta^{2}+\sin^{2}\theta \,d\phi^{2}
\right),
\end{equation}
where $B_0$ is a constant and $l=0.000001$ is
the tangential velocity.  So $\nu(r)=
\text{ln}\,B_0+l\,\,\text{ln}\,r $\,\,and
\begin{equation}\label{E:nuprime}
   \nu'(r)=\frac{l}{r}>0.
\end{equation}
On large scales, $\nu'(r)$ becomes negligible,
which is consistent with the FLRW model.

Returning once again to Eq. (\ref{E:lambda}),
we obtain next
\begin{equation*}
  \lambda'=\frac{1}{1+\frac{1}{4}K\,e^{\nu}\,(\nu')^2}
  \frac{1}{4}Ke^{\nu}[(\nu')^2+2\nu'\nu''],
\end{equation*}
and from Eqs. (\ref{E:E1}) and (\ref{E:E2}),
we have
\begin{multline}\label{E:F1}
   8\pi (\rho +3p)=e^{-\lambda}\left[
   \frac{1/r}{1+\frac{1}{4}K\,e^{\nu}\,(\nu')^2}
   \frac{1}{4}Ke^{\nu}[(\nu')^3+2\nu'\nu'']
   -\frac{1}{r^2}\right]\\+\frac{1}{r^2}
   +3e^{-\lambda}\left(\frac{1}{r^2}+
   \frac{\nu'}{r}\right)-\frac{3}{r^2}.
\end{multline}
Finally, it follows from Eq. (\ref{E:lambda})
that Eq. (\ref{E:F1}) can be written as
\begin{multline}\label{E:F2}
   8\pi (\rho +3p)=-\frac{2}{r^2}+
   \frac{2e^{-\lambda}}{r^2}+e^{-\lambda}
   \left[\frac{1/r}{1+\frac{1}{4}K\,
   e^{\nu}\,(\nu')^2}\frac{1}{4}Ke^{\nu}
   [(\nu')^3+2\nu'\nu'']+\frac{3\nu'}{r}
   \right]\\=-\frac{2}{r^2}
      \frac{\frac{1}{4}Ke^{\nu}(\nu')^2}
      {1+\frac{1}{4}K\,e^{\nu}\,(\nu')^2}
   +\frac{1}{1+\frac{1}{4}K\,e^{\nu}\,(\nu')^2}
   \left[\frac{1/r}{1+\frac{1}{4}K\,
   e^{\nu}\,(\nu')^2}\frac{1}{4}Ke^{\nu}
   [(\nu')^3+2\nu'\nu'']+\frac{3\nu'}{r}
   \right]\\
   =\frac{1}{1+\frac{1}{4}K\,e^{\nu}\,(\nu')^2}
   \left[-\frac{2}{r^2}\cdot\frac{1}{4}
   Ke^{\nu}(\nu')^2+\frac{1/r}
   {1+\frac{1}{4}K\,e^{\nu}\,(\nu')^2}
   \cdot\frac{1}{4}Ke^{\nu}
   \left[(\nu')^3+2\nu'\nu''\right]
   +\frac{3\nu'}{r} \right].
\end{multline}

The question now arises: what happens
when $K(t)$ passes zero to become
negative?  We already know that if $K$
is replaced by $-K$, Eq. (\ref{E:lambda})
becomes Eq. (\ref{E:lambda2}), while
Eq. (\ref{E:F2}) becomes
\begin{multline}\label{E:F3}
   8\pi (\rho +3p)=\\
    \frac{1}{1-\frac{1}{4}K\,e^{\nu}\,(\nu')^2}
   \left[\frac{2}{r^2}\cdot\frac{1}{4}
   Ke^{\nu}(\nu')^2-\frac{1/r}
   {1-\frac{1}{4}K\,e^{\nu}\,(\nu')^2}
   \cdot\frac{1}{4}Ke^{\nu}
   \left[(\nu')^3+2\nu'\nu''\right]
   +\frac{3\nu'}{r} \right], \\K>0.
\end{multline}
This result could also be obtained
directly from Eq. (\ref{E:lambda2}).

In view of Eq. (\ref{E:nuprime}), both
$(\nu')^3+2\nu'\nu''$ and $3\nu'/r$ are
negligibly small on large scales.  So
for $K(t)>0$, Eq. (\ref{E:F2}) implies that
$8\pi(\rho +3p)<0$.  This shows that
$\overset{..}{a}(t)$ in the Friedmann
equation is positive, indicating an
accelerated expansion.  For $K(t)<0$,
we obtain from Eq. (\ref{E:F3}) that
$8\pi(\rho +3p)>0$ and
$\overset{..}{a}(t)<0$, indicating a
decelerating expansion.

The deceleration will continue until
$K(t)=0$ to start a new cycle with
$\overset{..}{a}(t)>0$.  At this point,
the Universe would experience a big
bounce or a big-bang singularity.  For
a discussion of the big bounce, see
Poplawski \cite{nP16} and references
therein.  For the case of a singularity,
it needs to be stressed that the
embedding theory goes well beyond the
usual models in the following sense:
during the decelerating phase, we have
an extra timelike dimension.  This is
similar to Hawking's imaginary time
discussed in Ref. \cite{HH83}: while
ordinary time would still have a
big-bang singularity, imaginary time
avoids the singularity, also called
the no-boundary proposal.  Our extra
timelike dimension would have the
same effect, thereby allowing the
oscillations to continue unhindered.

Having a plausible mathematical model
does not automatically yield the best
physical interpretation, especially if
the model contains a free parameter.
So let us recall from Sec. \ref{S:embedding}
that $K$ started off as an arbitrary
constant that was subsequently replaced
by a periodic function of $t$ without
affecting the solution.  The best physical
interpretation of the resulting model
appears to be the signature change
discussed in Sec. \ref{S:K(t)} simply
because the outcome complements the
various models in Refs. \cite{WWV10,
STP18, LMT21,  aU08, PPG, MSV08, BB20}.

\section{Conclusions}\label{S:summary}
This paper begins with a brief review of
the literature dealing with
signature-changing events in various
gravitational theories.  This paper
discusses these issues in the context of
the well-established embedding theory
in an $n$-dimensional Riemannian space.
Such a space is said to be of embedding
class $m$ if $m+n$ is the lowest dimension
$d$ of the flat space in which the given space
can be embedded; here $d=\frac{1}{2}n(n-1)$.
So a four-dimensional Riemannian space
is of class two since it can be embedded 
in a six-dimensional flat space.  We
also made use of the fact that a metric
of class two can be reduced to a metric of
class one by a suitable coordinate
transformation.  Moreover, Einstein's
theory allows the extra dimension to be
either spacelike or timelike.

It is first shown that a spacelike extra
dimension leads to an accelerated
expansion, thereby accounting for the
mysterious dark energy.  An extra timelike
dimension, on the other hand, leads to a
decelerating expansion.  Our mathematical
model includes a free parameter $K$ which we
can take to be a periodic function of time,
resulting in an oscillating solution.
This assumption can be justified for physical
reasons: we have seen that the best physical
interpretation comes from the well-established
embedding theory that clearly points to a
signature-changing event.  This outcome is
is very much in line with the various models
in Refs. \cite{WWV10, STP18,
LMT21,  aU08, PPG, MSV08, BB20}.
An important additional feature of the
embedding space is the extra timelike
dimension during the decelerating phase.
This allows the oscillations to continue
whether the transition involves a big-bang
singularity or just a big bounce.
\\
\\\noindent
\textbf{Conflicts of interest}\\
The author declares no conflicts of interest
regarding the publication of this paper.

\end{document}